\documentclass[aps,prl,floats,twocolumn,floatfix,showpacs]{revtex4}
\usepackage{psfig,epsf,graphicx}
\usepackage{amssymb}
\usepackage{amsmath}
\usepackage{eepic}

\newlength{\piclen}
\begin{document}

\title{
Localization of strongly  correlated electrons as Jahn-Teller polarons in  manganites}

\author{Y.-F.\ Yang and  K.\ Held}
\affiliation{Max-Planck-Institut f\"ur
Festk\"orperforschung, 70569 Stuttgart, Germany}

 \date{March 21, 2006}

\begin{abstract} 
A realistic modeling of manganites
should include
the Coulomb repulsion between $e_g$
electrons, the Hund's rule coupling to $t_{2g}$ spins,
and Jahn-Teller phonons.
Solving such a model  by dynamical mean field theory,
we report 
large magnetoresistances and
spectra in good agreement with experiments.
The physics  of the unusual, insulating-like
paramagnetic phase is determined 
by correlated electrons which
are---due to strong correlations---easily trapped 
as   Jahn-Teller polarons.
\end{abstract}

% 71.27.+a  Strongly correlated electron systems; heavy fermions
% 71.10.Fd  Lattice fermion models (Hubbard model, etc.)
% 71.30.+h  Metal-insulator transitions and other electronic transitions
% 71.20.Eh  Rare earth metals and alloys
% 75.20.Hr  Local moment in compounds and alloys; Kondo effect, valence
%           fluctuations, heavy fermions
% 75.47.Gk  Colossal magnetoresistance

\maketitle

Since large magnetoresistances are 
of substantial technological importance,
the colossal magnetoresistance (CMR) \cite{CMR}
in
manganites  (T$_{1-x}$D$_x$MnO$_3$;
T: trivalent ion, e.g., La;
D: divalent ion, e.g., Ca)
has attracted considerable
attention, both experimentally and theoretically 
\cite{reviews}.
Like the mechanism for high-temperature superconductivity in the
cuprates, the CMR still lacks a thorough theoretical understanding;
and like superconducting transition temperatures,
the temperatures for the CMR are much lower than technologically  desirable.

The key to the understanding of CMR lies in the unusual properties
of the paramagnetic (PM) phase which---in a wide range of dopings $x$---shows
an unusual insulating-like behavior.
At lower temperatures $T$ and/or
within magnetic fields,  the double exchange mechanism \cite{Zener51a}
stabilizes a ferromagnetic (FM) metallic phase.
If the insulating-like PM phase was understood, a CMR could simply be described
 as the PM-insulator-to-FM-metal transition.
Different proposals aim at explaining the CMR
and the PM insulating-like phase,
in particular,
 the localization of
charge carriers through lattice \cite{Latticepolarons}
 or orbital  \cite{Kilian99b} polarons, 
an Anderson localization arising from
disordered $t_{2g}$-spins \cite{Varma96a},
a phase separation into nano-domains showing percolation effects
\cite{Dagotto1,Dagotto}, and an effective
 two band model \cite{Ramakrishnan04}.
%af spin fluctuations?
However, hitherto no quantitative, microscopic calculation
satisfactorily explains
 the known physical properties of the PM
insulating-like phase, which is not only characterized by
an increase of the resistivity $\rho$ with decreasing $T$,
but also has unusual dynamic 
properties. The latter reflect in a
spectral function $A(\omega)$ with a very low
spectral weight
at the Fermi level $E_F$ irrespectively of $x$,
as indicated by photoemission and X-ray absorption 
experiments  \cite{PES,Park}.
Similarly the optical conductivity $\sigma(\omega)$ shows a very low spectral
weight up to an energy scale of $\sim\! 1\,$eV \cite{optics}.
Also the FM is an atypical (bad) metal.

The theoretical understanding is complicated since
several ingredients seem to be necessary to
describe manganites:
the exchange interaction between more localized $t_{2g}$ spins
and the itinerant $e_g$ electrons \cite{Zener51a},
the Jahn-Teller phonons and their coupling to the electrons \cite{Latticepolarons},
as well as the electronic correlations due to the  local
Coulomb interaction \cite{Imada98a,Held00}.
The necessity of all three of these  interactions
was also revealed by realistic
 LDA+DMFT (local density approximation + 
dynamical mean field theory \cite{LDADMFT}) calculations
for the parent compound LaMnO$_3$,
showing 
the  metal-insulator 
transition in LaMnO$_3$ under pressure \cite{Loa}
to be triggered by the (Jahn-Teller) crystal field splitting
 which is strongly enhanced
by the local Coulomb interaction \cite{Yamasaki06}.

In this Letter, we model doped  LaMnO$_3$
by including a realistic tight-binding band
structure for cubic LaMnO$_3$,
the Coulomb interaction between the $e_g$ electrons, Jahn-Teller
phonons, and the Hund's exchange coupling to localized $t_{2g}$ spins.
We solve this model by dynamical mean field theory  (DMFT) \cite{DMFT},
employing the (numerically) exact quantum Monte Carlo (QMC) \cite{QMC}
method as an impurity solver.
We find that  tendencies
of the  Coulomb repulsion
and the Jahn-Teller distortion
to localize electrons mutually
support each other, trapping
$e_g$ electrons as local polarons in the PM phase. 
Hence the PM phase is insulating-like with
very low optical spectral weight
below 1eV and  with a resistivity which 
is by a factor of $\sim\! 8$
larger than that of the FM phase (magnetoresistance).
Jahn-Teller polarons
without correlation effects have been studied
before, e.g.\ in 
\cite{Latticepolarons,Millis96,Michaelis03},
but suffered from   shortcomings such as
big magnetoresistances only for  undoped 
LaMnO$_3$  (which is always insulating in experiments)
and discrepancies with experimental spectra.

{\it Model for manganites.}
Starting point of our investigation is the following, realistic model
for manganites
  \begin{eqnarray}
    \hat{H}&=& - \sum_{l,m=1}^{2} \sum_{\langle i j \rangle \sigma} t^{i j}_{lm}
  {\hat{c}}^{\dagger}_{{i} l\sigma}   {\hat{c}}^{\phantom{\dagger}}_{{j} m\sigma} 
   - 2  {\cal J} \;     \sum_{m i} {\hat{\bf  s}}_{i m}   \;
   {\hat{\bf  S}}_{i}\nonumber \\&&
    \!\!+   { U}   \sum_{m i} 
    \hat{n}_{im\uparrow}\hat{n}_{im\downarrow}
     + \sum_{i \,  \sigma \tilde{\sigma}} 
    ( U'\!-\!\delta_{\sigma \tilde{\sigma}}J)  \;
    \hat{n}_{i  1 \sigma} \hat{n}_{i 2  \tilde{\sigma}}\nonumber \\&&
\!\!+g\sum_{i\sigma}\sum_{m_1,m_2=1}^{2}\hat{c}^{{\dagger}}_{i m_1 \sigma}(\hat{Q}_2{\bf \;
  \tau}^z+\hat{Q}_3{\bf \tau}^x)_{m_1m_2}
\hat{c}^{\phantom{\dagger}}_{i m_2 \sigma}\nonumber \\&&
  \!\!+ \sum_{a=2}^{3}(\frac{1}{2} \hat{P}_a^2 + \frac{1}{2} \Omega^2 \hat{Q}_a^2)
    \label{Eq:Ham}
  \end{eqnarray} 
Here, 
 ${\hat{c}}^{\dagger}_{i m \sigma}$ and
${\hat{c}}^{\phantom{\dagger}}_{i m \sigma}$
are creation and annihilation operators for electrons
on site $i$ within $e_g$ orbital $m$ and 
spin $\sigma$; 
${\hat{\bf s}}_{i \nu}= \frac{1}{2}\sum_{\sigma \sigma'}  
 \hat{c}^{\dagger}_{i \nu \sigma}  
 {\mathbf \tau}^{\phantom{\dagger}}_{\sigma \sigma'}
 \hat{c}^{\phantom{\dagger}}_{i  \nu \sigma'}$  denotes the e$_{{g}}$-spin
(${\bf \tau}$: Pauli matrices), ${\hat{\bf  S}}_{i}$ the t$_{{2g}}$-spin, and $\hat{Q}_a$ ($\hat{P}_a$) the  coordinate (momentum)
of the two quantum Jahn-Teller phonons.
Let us briefly discuss the three lines of
Hamiltonian (\ref{Eq:Ham}).
The  first line describes the Kondo
lattice model
 consisting of two terms:
(i) the tight binding band structure 
with directionally dependent nearest-neighbor
hopping,
$t^x_{11}\!\!=\!\!-\sqrt{3}t^x_{12}\!\!=\!\!-\sqrt{3}t^x_{21}\!\!=\!\!3t^x_{22}\!\!=\!\!3t_0/4$,
 $t^y_{11}\!\!=\!\!\sqrt{3}t^y_{12}\!\!=\!\!\sqrt{3}t^y_{21}\!\!=\!\!3t^y_{22}\!\!=\!\!3t_0/4$,
 $t^z_{22}\!\!=\!\!t_0$, and
 $t^z_{11}\!\!=\!\!t^z_{12}\!\!=\!\!t^z_{21}\!\!=\!\!0$ with bandwidth 
 $W\!\!=\!\!6t_0\!\!=\!\!3.6\,$eV which well describes the LDA
band structure for the cubic lattice 
(contributions of
longer-range hopping  are very minor
  \cite{Yamasaki06b}); and
(ii) the coupling  ${\cal J}$ to  the t$_{{2g}}$ spin
which we assume to be classical with
strength $2 {\cal J} |{\hat{\bf  S}}_{i}|=2.66$ calculated
from a  ferromagnetic LDA calculation  \cite{Yamasaki06}.
The Kondo lattice model (first line) 
can be solved exactly in DMFT
\cite{Furukawa94a} and gives rise to the
double exchange mechanism \cite{Zener51a}
for ferromagnetism.
The second line describes the Coulomb interactions
between the $e_g$ electrons including
the intra-  ($U$) and inter-orbital repulsion $U'$,
as well as the Hund's exchange ($J$).
Even for large  ${\cal J}$, the Coulomb repulsion
$U'-J$ between two spin-aligned $e_g$ electrons on the
same lattice site is important.
It leads to the formation of Hubbard bands and
quasiparticle peaks in the PM phase
 \cite{Held00}.
We take $U'=3.5\,$eV and $J=0.75\,$eV from Ref.\ \cite{Park}
($U=U'+2J$ follows by symmetry).
Finally, the third line describes the coupling $g$
of the $e_g$ electrons to the two 
local Jahn-Teller phonons;
and the fourth line the quantum motion of these phonons.
The phonon frequency $\Omega=0.07\,$eV is estimated from Raman
spectra and the results of lattice dynamical calculations \cite{Iliev}; the
 breathing mode, which has a higher frequency and
is believed to be less important \cite{Dagotto},
is neglected.
Since the electron-phonon coupling is difficult
to determine from band structure data,
we consider different values, mostly
$g\!=\!0.10\,$eV$^{3/2}$ which corresponds to  a  
dimensionless coupling constant
$\lambda\!=\!g/\Omega\sqrt{t}\!=\! 1.84$,
comparable---slightly larger---than 
previous estimates, see \cite{Dagotto}.
In the following, we set $\hbar\!=\!e\!=\!k_B\!=\!1$;
our unit of energy is eV.

{\it DMFT(QMC) implementation.}
For solving  Hamiltonian (\ref{Eq:Ham}) by DMFT(QMC), the
imaginary time $\tau\!=\!0..\beta (\equiv 1/T)$ is discretized
into $l\!=\!1..L$ Trotter slices $\tau_l$ of 
width $\Delta\tau\!=\!0.2$, and
the Coulomb interactions are decoupled
through discrete Hubbard-Stratonovich
transformations.
We sample the auxiliary Hubbard-Stratonovich field
and the continuous  phonon Bose field $Q_a(\tau_l)$ 
on equal footing, after replacing
the phonon momentum $P_a(\tau_l)$
by $[Q_a(\tau_l)-Q_a(\tau_{l-1})]/\Delta \tau$
as discussed in \cite{BSS2}.
After each QMC sweep, a global  update of the 
phonon field
 is considered; we use 
$\sim\!10^6\!-\!10^7$ sweeps.

\begin{figure}[t]
{\includegraphics[width=5.3cm,angle=270]{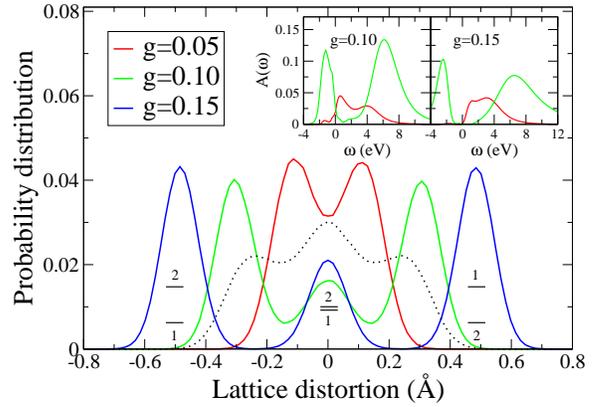}}

\caption{(Color online)
{QMC distribution of the lattice distortion
[$1/L \sum_{l=1}^L Q_2(\tau_l)$] for different
electron-phonon couplings $g$, $n\!=\!0.8$ electrons/site and $\beta\!=\!16$;
the dotted line is
  without Coulomb repulsion  for comparion ($g\!=\!0.1$).
At strong enough $g$, three regions with large (positive and negative)
and small lattice distortion emerge.
The inset shows the electron spectral function 
for these regions with
large (light/green) and small (dark/red) lattice distortions, respectively. 
We can interpret the results physically as the trapping
of the $n=0.8$ electrons  as
Jahn-Teller polarons. 
\label{Fig:Polaron}}}
\end{figure}

{\it Polaron formation in the PM phase.}
Let  us start our discussion  with  
the QMC distribution of the $\tau$-averaged phonon field $Q_2$,
  corresponding
to the amplitude of the lattice distortion \cite{footnoteunits}.
Fig.\  \ref{Fig:Polaron} shows three peaks
with large (positive and negative) Jahn-Teller distortion
and small  Jahn-Teller distortion, respectively.
Upon lowering temperature, and at strong enough
electron-phonon coupling, 
these three peaks get sharper and clearly separated in the PM phase.
Then, the QMC sampling tends to stay within one of the
three regions for a very long computational time \cite{note1}.

\begin{figure}[t]
{\includegraphics[width=5.3cm,angle=270]{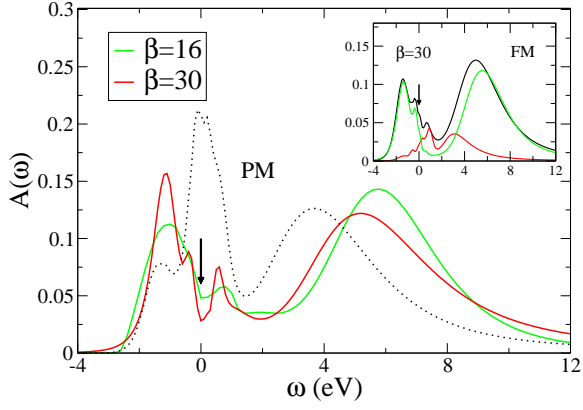}}

\caption{(Color online)
{PM spectrum as a function of temperature for $g=0.1$ and $n=0.8$.
With decreasing $T$, the polaron formation is stabilized and
 a pseudo-gap opens at $E_F\!=\!0$.
The dotted line is the PM  spectrum without electron
phonon coupling at $n=0.8$, $\beta=30$,
showing a  damped and renormalized
quasiparticle peak at $E_F$.
Inset: total FM spectrum in the presence
 (light/green) and absence (dark/red) of 
lattice distortion, analogous to 
the inset of Fig.\ \ref{Fig:Polaron} for the PM phase.
Due to the enhanced mobility in the FM phase,
the Jahn-Teller distorted band (i.e., the 
lower light/green band) 
is wider, not fully occupied, and hence metallic-like.
\label{Fig:PMspectrum}}}
\end{figure}

Physically, the
three sharp peaks
mean that the local lattice site
is either strongly Jahn-Teller distorted 
(with orbital 1 or 2 lower in energy as visualized in Fig.\ 
\ref{Fig:Polaron}) 
or not (central peak).
This clear separation into three peaks furthermore suggests 
that the phonon 
dynamics is strongly reduced.
In our view, the 
slow QMC dynamics for going from 
one of the two distorted to the undistorted
configurations indeed reflects a slow
real time dynamics \cite{note2}.
Turning to the electronic spectrum
in the inset of Fig.\  \ref{Fig:Polaron},
we see there is 
one electron with Jahn-Teller distortion
(the peak  below $E_F=0$ of the light/green line
contains one electron)
and almost none without distortion 
(the dark/red line is above the Fermi energy).
This is exactly the picture of a local polaron, 
an electron trapped through the lattice distortion.
This trapping  explains why we obtain 
an insulating-like spectrum at  $g\!=\!0.1$
and even a true gap for  $g\!=\!0.15$.
Let us note that the polaron formation is strongly 
supported by the local Coulomb repulsion.
Without  $U'$,
the polaron formation fades away (dotted line
in Fig.\   \ref{Fig:Polaron}). The effect of
$U'$
 on the polaron formation is two-fold:
(i) It strongly enhances the Jahn-Teller splitting 
from $2E_{\rm JT}\!\!=\!\!g^2/\Omega^2$ to  
$2E_{\rm JT}+U'-J$, see the (light/green) spectrum
in the presence of distortion in the inset of Fig.\   \ref{Fig:Polaron}.
Since this splitting has to overcome the $e_g$ bandwidth 
for a polaron localization, $U'$ is actually 
needed---without $U'$ no strong lattice distortion 
emerges (dotted line).
(ii) Due to strong quasiparticle renormalizations
the effective bandwidth of the correlated bands is 
strongly reduced
(see the dotted line in Fig.\ \ref{Fig:PMspectrum}).
We can estimate  this bandwidth 
renormalization to be roughly $Z=(1-\partial {\rm Re} \Sigma(\omega)/\partial\omega|_{\omega=0})^{-1}\sim 0.4$,
as obtained from  the self energy
at real $\omega$.
Note, however, that these quasiparticles are strongly damped 
through  scattering at $t_{2g}$ spins
 (${\rm Im}\Sigma(0)\sim -0.7\,$eV).
Since the correlated electrons are already more localized,
it is easy to trap them as polarons.

{\it Spectrum in the PM phase.}
Let us turn to the total (distortion-averaged) 
PM spectrum  in Fig.\ \ref{Fig:PMspectrum}.
It shows,  upon decreasing $T$, 
 the development of a pseudogap at $E_F$
which 
separates  localized Jahn-Teller 
polarons and undistorted configurations
without electrons.
This  pseudogap is more 
enhanced at lower $T$ since---without 
thermal smearing of the lattice distortion---the 
three peaks in Fig.\   \ref{Fig:Polaron}
become sharper. Consequently,
there is a sharper separation (pseudogap)
between
polaron trapped electrons and 
undistorted unoccupied states.

\begin{figure}[t]
{\includegraphics[width=5.3cm,angle=270]{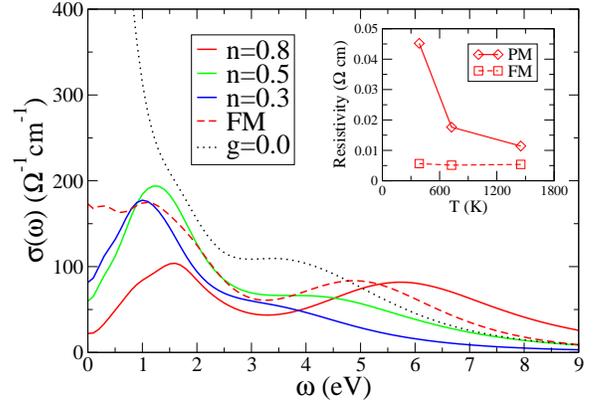}}
\caption{(Color online)
{Optical conductivity $\sigma(\omega)$
for the PM phase at $g\!=\!0.1$, $\beta\!=\!30$ and $n\!=\!0.8$, 0.5, and 0.3 electrons/site,
showing an insulating-like behavior with a
 reduced weight below 1eV.
The dotted line shows a metallic Drude peak 
in the absence of the electron-phonon coupling ($\beta\!=\!30$, $n\!=\!0.8$);
the dashed line  the bad metallic behavior of the FM phase 
($g\!=\!0.1$, $\beta\!=\!30$, $n\!=\!0.8$). 
Inset: The PM resistivity strongly increases with decreasing $T$ (here for
$g\!=\!0.1$ and $n\!=\!0.8$)
so that a transition to the FM phase reduces the resistivity by a 
factor of $\sim\! 8$. There is a ``colossal'' magnetoresistance.}
\label{Fig:CMR}}
\end{figure}

In Fig.\ \ref{Fig:CMR}, we see that
the pseudogap of the PM spectrum  leads
to a similar pseudogap  in the optical spectrum 
at small frequencies, so that the optical spectral 
weight is very
much reduced below $1$eV \cite{footnoteunits}.
The first optical peak at $\sim \! 1.5$eV  (named midgap state in the literature)
then stems from exciting the polarons into the unoccupied,
undistorted band above the Fermi energy.
At  $\sim 5\!-\!6\,$eV, a second peak
in the optical spectrum signals 
the excitations of states with two
$e_g$ electrons per site, which correspond
to the upper (light/green) peak in the inset of Fig.\ \ref{Fig:Polaron}.
The reduced optical spectral weight at low frequencies,
and (approximately) the position
of the two optical peaks are in agreement with experiment 
\cite{optics}, as is the  shift of the two
peaks towards lower frequencies with increasing hole doping 
$x$ (decreasing $n\!=\!1\!-\!x$) and the gradual filling of 
the optical
gap with increasing $x$ \cite{optics}.
Let us note 
that the height of
the two optical peaks does not fully agree with experiment:
In contrast to Fig.\ \ref{Fig:CMR},
 the
second peak has experimentally
several times more spectral weight
than the first (midgap) peak.
We can understand this disagreement
since there are  
other 
optical contributions in this energy range:
from  oxygen charge transfer, from La bands,
and the  upper Hubbard band 
of the $t_{2g}$ electrons. 
Such effects are beyond our
low-energy Hamiltonian (\ref{Eq:Ham}).
Also the first (midgap) peak has experimentally
 a larger $\sigma(\omega)$ than in Fig.\ \ref{Fig:CMR};
for a smaller $g$,
$\sigma(\omega)$ would increase  and the two peaks would
shift to lower $\omega$.

From $\sigma(0)$, we directly obtain the resistivity
in the inset of Fig.\  \ref{Fig:CMR}.
As the electrons are more strongly
 trapped as Jahn-Teller polarons
with decreasing temperature, the resistivity
of the PM phase is strongly enhanced.
We observe an insulating-like behavior.
%%%
%

{\it (Bad) metallic FM phase.}
Let us finally turn to the FM phase which we
simulate by fixing the direction of the $t_{2g}$ spins.
The FM spectrum for $n=0.8$ and $\beta=30$
is presented in the inset of Fig.\ \ref{Fig:PMspectrum},
separated into contributions from small and large Jahn-Teller distortion.
While the peak
positions of the distorted and undistorted
bands are at similar energies as for the PM
in Fig.\ \ref{Fig:Polaron}, the widths of these bands
are much larger so that the band with Jahn-Teller distortion
crosses the Fermi level and is hence metallic-like.
This is also reflected in the self-energy
which, in contrast to the PM phase, 
behaves metallic-like 
with $\partial {\rm Re} \Sigma (\omega)/ \partial\omega<0$,
corresponding to a quasiparticle weight of 
roughly $Z\!\sim\!0.6$. However,
these quasiparticle excitations are very strongly damped
since, when moving through the crystal, they
scatter at different local potentials, with and without
Jahn-Teller distortion. 
The  scattering rate is given by ${\rm Im} 
\Sigma(0)\sim -1.9\,$eV which corresponds to
a  quasiparticle life time  of half a femto second. It is hence questionable whether we
can still speak of quasiparticles at all. In any case, a 
bad metallic behavior is to be expected and indeed seen
in the optical conductivity \cite{optics}, showing 
a wide peak at low frequencies 
(dashed line in Fig.\ \ref{Fig:CMR}).

Having described the insulating-like PM with a pseudogap
in the spectrum and the FM without this pseudogap,
 the magnetoresistance simply 
follows from the PM-FM transition---triggered by temperature
or magnetic fields. 
For $n=0.8$, the inset of 
 Fig.\ \ref{Fig:CMR} yields
a change of resistivity  by about a factor 8
when going from the PM to the FM phase.

{\it Conclusion.}
We reported how the tendencies of
Coulomb repulsion
and Jahn-Teller distortion
to localize electrons mutually 
enhance each other
so that correlated electrons are trapped as polarons
in the PM phase of doped manganites.
In the FM phase, on the other hand,
electrons are still mobile, but strongly 
scattered because of fluctuating (strong and weak)
Jahn-Teller distortions, yielding 
a bad metal.
The transition from PM to FM 
is hence accompanied by a
``colossal'' magnetoresistance.

We thank
O.\ K.\ Andersen, N.\ Bl\"umer, N.\ N.\ Kovaleva, 
O.\ R\"osch,
A.\ Yamasaki,
and particularly J.\ E.\ Han, who explained to us
 how he treats phonons in QMC,
for discussions; and  the Deutsche Forschungsgemeinschaft
for financial support
 through the Emmy Noether program.


\begin{thebibliography}{99}
\bibitem{footnoteunits}
We use the  mass  of the oxygen atom
for getting the lattice distortion in \AA { }
in Fig.\ \ref{Fig:Polaron};
and the lattice constant $a\!=\!5.5\,$\AA { }
 for  getting the optical conductivity 
in Fig.\ \ref{Fig:CMR}.
\bibitem{note1} In the extreme
cases ($g=0.15$; $\beta=16$) and
 ($g=0.1$; $\beta=30$),
it is easy to QMC sample the regions of the  
three different peaks independently.
But because of the extremely low propability to 
go from one region to another, the relative importance
of the three well-separated regions
is not adequately sampled by QMC.
This relative weight can, on the other hand,
be extrapolated from high $T$, which agrees
with the physical expectation that the
 weight is  $1-n$ and
 $n/2$ 
for small distortion and the two regions with large distortions,
respectively. For the final results for the above-mentioned cases, 
we hence weight the independently sampled regions accordingly.
\bibitem{note2}
Note that a very similar (two maxima)
distribution of the averaged Ising-like field is 
observed for the DMFT Mott insulator, where
it indicates the presence of a localized spin (up or down);
and the slow QMC dynamics from spin up to down 
reflects the long-time stability of the spins
in the Mott-insulating phase of the Hubbard model.
\end{thebibliography}
\end{document}